# Nonvolatile Tuning of Bragg Structures Using Transparent Phase-Change Materials

NICHOLAS A. NOBILE,[1,†] CHUANYU LIAN,[2,3,†] HONGYI SUN,[2,3] YI-SIOU HUANG,[2,3] BRIAN MILLS,[4,5] COSMIN CONSTANTIN POPESCU,[4] DENNIS CALLAHAN,[6] JUEJUN HU,[4] CARLOS A. RÍOS OCAMPO,[2,3] AND NATHAN YOUNGBLOOD[1,*]

[1]*Electrical & Computer Engineering Department, The University of Pittsburgh, Pittsburgh, PA 15213, USA*
[2]*Department of Materials Science & Engineering, University of Maryland, College Park, MD 20742, USA*
[3]*Institute for Research in Electronics & Applied Physics, University of Maryland, College Park, MD 20742, USA*
[4]*Department for Materials Science & Engineering, Massachusetts Institute of Technology, Cambridge, MA 02139, USA*
[5]*Draper Scholar Program, The Charles Stark Draper Laboratory, Inc, Cambridge, MA, 02139*
[6]*The Charles Stark Draper Laboratory, Inc, Cambridge, MA, 02139*
[†]*These authors contributed equally*
*nathan.youngblood@pitt.edu*

**Abstract:** Bragg gratings offer high-performance filtering and routing of light on-chip through a periodic modulation of a waveguide's effective refractive index. Here, we model and experimentally demonstrate the use of $Sb_2Se_3$, a nonvolatile and transparent phase-change material, to tune the resonance conditions in two devices which leverage periodic Bragg gratings—a stopband filter and Fabry-Perot cavity. Through simulations, we show that similar refractive indices between silicon and amorphous $Sb_2Se_3$ can be used to induce broadband transparency, while the crystalline state can enhance the index contrast in these Bragg devices. Our experimental results show the promise and limitations of this design approach and highlight specific fabrication challenges which need to be addressed in future implementations.

## 1. Introduction

Wavelength division multiplexing (WDM) with frequency selective routing, filtering, and modulation is one of the core advantages of optics over electronics for data transmission. The ability to control several THz of bandwidth in the telecommunications bands has transformed the way data is sent locally within data centers and globally in transatlantic fiber communications. Frequency selective control of light on-chip is equally important for a variety of commercial and emerging applications, such as on-chip laser cavities [1], [2], resonant modulators [3], pulse shaping [4], LiDAR [5], [6], and even computing [7]–[10]. One simple, yet powerful technique to control the wavelength-dependent response of an optical signal is through Bragg gratings which are typically fabricated using periodic perturbations to the waveguide width [11]. By changing the period and modulating the strength of these perturbations, the bandwidth and central frequency of the Bragg grating can be controlled [12]. Additionally, filters comprised of Bragg gratings are not limited in channel density by free spectral range (FSR) effects which are an issue for microring-based WDM filter banks [13].

Due to the fixed nature of geometric patterning, the tunability of Bragg gratings is limited after fabrication. Thermo-optic or electro-optic effects can be used to tune the resonance condition within a limited range [3], [14], [15], although they are volatile and require constant power supply to tune the device. Creating reconfigurable, nonvolatile photonic filters could simplify system design and enable multi-functionality within the same circuit on-chip. Additionally, the grating profile on-chip can be subject to fabrication variations and nonvolatile methods for tuning the resonance frequency of these Bragg gratings could be important for aligning multiple FP resonators or contra-directional couplers on-chip [3], [8]. Low-loss phase-change materials are ideal for this application as the high index contrast between the amorphous and crystalline phases can be used to create a periodic index perturbation or tune the resonance of a Bragg grating [16], [17]. While prior results have demonstrated this concept experimentally using $Ge_2Sb_2Te_5$, the high absorption in the crystalline state limits the spectral performance of these devices and increases insertion loss [18], [19].

Here, we propose and experimentally explore the ability to switch between enabling and disabling the Bragg resonance within a periodic device that is functionalized with the phase-change material $Sb_2Se_3$. Through eigenmode simulations and transfer matrix method (TMM) modeling of these devices, we show that it is possible to use the large and transparent index contrast of $Sb_2Se_3$ to either amplify or cancel out the effective index contrast of a Bragg grating by carefully designing the width of the waveguide with and without $Sb_2Se_3$. Additionally, we show that the large index contrast between the amorphous and crystalline states can be used to tune the Fabry-Perot (FP) resonance of a phase-shifted Bragg grating and even completely move it beyond the stopband of the grating under certain design conditions.

## 2. Designing tunable Bragg gratings with phase-change materials

Fig. 1a illustrates the overall concept of our design approach. A Bragg grating with $Sb_2Se_3$ embedded in the waveguide (dark blue and red regions) can either cancel or enhance the periodic perturbation of the effective refractive index of the waveguide. When the effective refractive index of waveguide with amorphous $Sb_2Se_3$ matches that of the silicon waveguide without $Sb_2Se_3$, the effective index perturbation is canceled, and the device appears to the propagating optical mode as if no grating exists. However, after crystallization, the regions with embedded $Sb_2Se_3$ further enhance the index contrast and perturbation from the patterned Bragg grating is increased. This effect is directly related to fact that in the amorphous state, $Sb_2Se_3$ has a slightly lower refractive index than silicon ($n_{am} = 3.27$ at $\lambda = 1550$ nm) while in the crystalline state the refractive index is higher than that of silicon ($n_{cry} = 4.04$ at $\lambda = 1550$ nm). The refractive indices of $Sb_2Se_3$ thin films before and after crystallization were measured using ellipsometry and are shown in Fig. 1b.

We explored two design approaches to control the index perturbation of the Bragg gratings. The first design illustrated in Fig. 1c embeds the $Sb_2Se_3$ directly in the waveguide, replacing silicon in the embedded regions. Using Ansys Lumerical's eigenmode solver (MODE), we simulated the effective refractive index of a waveguide with $Sb_2Se_3$ of various widths embedded 120 nm into a standard 500 nm × 220 nm single mode silicon waveguide. For increasing $Sb_2Se_3$ widths, the effective refractive index decreases in the amorphous state (due to a lower refractive index compared to that of bulk silicon) and increases in the crystalline state. If we choose to replace a 120 nm × 120 nm section of the silicon waveguide with $Sb_2Se_3$, $n_{eff}$ in the amorphous state is equivalent to that of a 480 nm × 220 nm silicon waveguide. Thus, patterning a Bragg grating in a 500-nm-wide silicon waveguide with a 20 nm wide sidewall perturbation and embedding amorphous $Sb_2Se_3$ in the 500-nm-wide sections results in a waveguide with no effective index perturbation and no Bragg resonance. Crystallization changes $n_{eff}$ significantly in the embedded regions with $\Delta n_{eff} \approx 0.165$ as shown in the righthand graph in Fig. 1d.

In addition to embedding the $Sb_2Se_3$ into the waveguide, we also considered a design with $Sb_2Se_3$ deposited directly on top of the waveguide as shown in Fig. 1e. This design is easier to fabricate but reduces the change in modulation strength of the Bragg grating to $\Delta n_{eff} \approx 0.074$ since the interaction between the PCM and the optical mode is reduced to evanescent coupling. Additionally, since material is being added to the waveguide (rather than replacing the silicon), the effective refractive index is increased for both the amorphous and crystalline phases. Therefore, for this evanescently-coupled design, $Sb_2Se_3$ is added to the regions of the waveguide that are narrower than the plain silicon waveguide in order to maintain a constant $n_{eff}$ in the amorphous phase. The conditions for effective index matching with a 520-nm-wide waveguide and the effective index change after crystallization are again shown in the righthand graph in Fig. 1f.

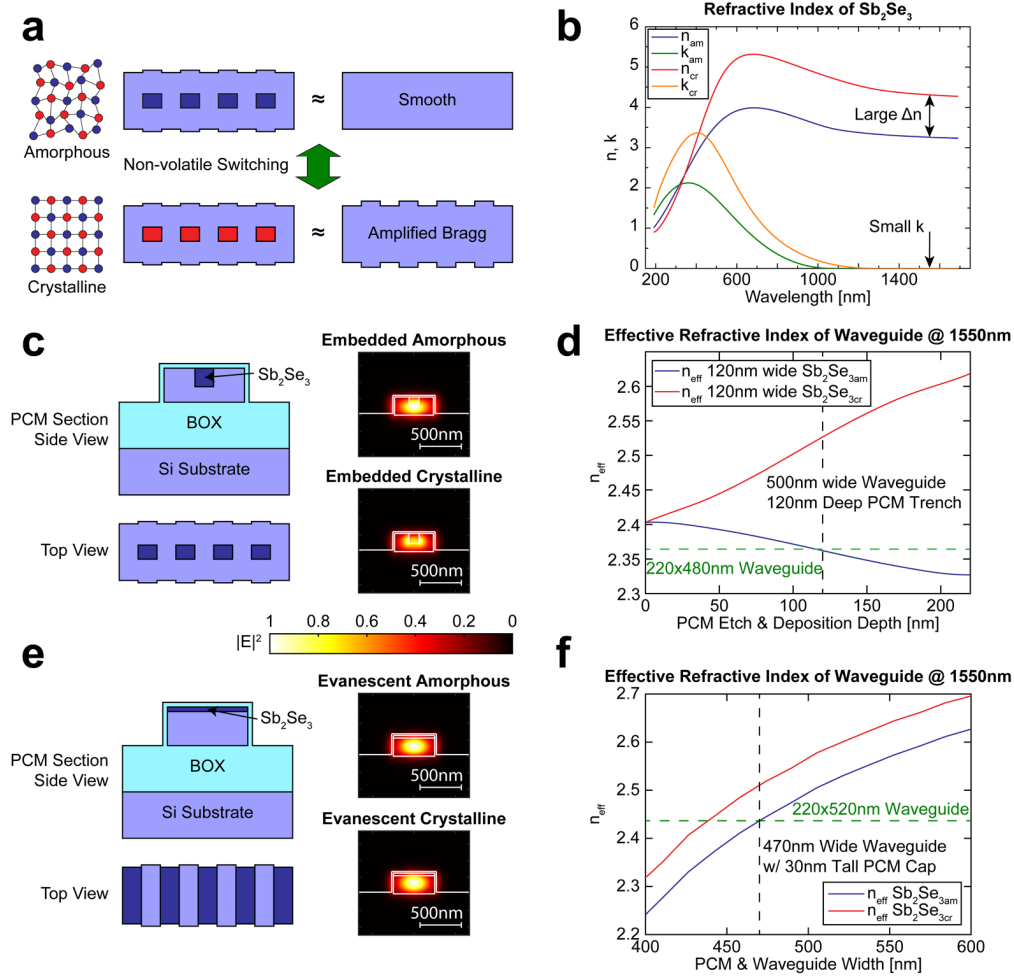

Fig. 1. (a) Proposed concept for switching between a resonant Bragg grating with enhanced index contrast (bottom) and broadband transmission with no periodic index contrast (top) using the nonvolatile phase transition of $Sb_2Se_3$. (b) Measured refractive index for as-deposited (amorphous) $Sb_2Se_3$ and annealed (crystalline) $Sb_2Se_3$ using thin-film ellipsometry. (c-f) Illustration of $Sb_2Se_3$ embedded in the waveguide (c-d) or deposited on top of the waveguide (e-f). The effective refractive index of the combined waveguide-$Sb_2Se_3$ system in both the crystalline and amorphous states is shown in the graphs on the right (c, e). The design conditions where the effective refractive index of the waveguide with amorphous $Sb_2Se_3$ matches that of a waveguide without any $Sb_2Se_3$ is indicated by the intersection of the dashed green and black lines in the graphs on the right (d, f).

## 3. Modeling Results

*Modeling Bragg gratings with $Sb_2Se_3$*

Using the $n_{eff}$ results from Fig. 1c-d, we used the transfer matrix method (TMM) to simulate the spectra of our two proposed designs [11]. This modeling approach uses the Fresnel Equation which approximates the Bragg grating as a periodic step perturbation in the waveguide's $n_{eff}$. These simulations account for the dispersion of the waveguide geometry and refractive index of the $Sb_2Se_3$ in both states. Fig. 2a-b show the resulting spectra for the embedded design, while Fig. 2c-d are for the evanescently coupled design for $N = 100$, where

$N$ is the number of periods. While both designs can achieve broad transmission close to unity over the simulated 1.5–1.6 µm wavelength range, we see that the stopband in the crystalline state is significantly narrower with a lower peak reflection for the evanescently coupled design (Fig. 2c). This is due to the weaker modulation of $\Delta n_{eff}$ compared to the embedded design which is directly related to the bandwidth of the Bragg grating [11]:

(1) $$\Delta\lambda = \frac{\lambda_B^2}{\pi n_g}\sqrt{\kappa^2 + (\pi/L)^2}, \text{ with } \lambda_B = 2\Lambda\bar{n}_{eff} \text{ and } \kappa = \frac{2\Delta n_{eff}}{\lambda_B},$$

where $\Delta\lambda$ is the bandwidth of the Bragg filter measured between the first nulls around resonance, $\lambda_B$ is the central resonance wavelength, $\Lambda$ is the grating period, $\bar{n}_{eff}$ is the average effective index of the grating, $n_g$ is the group index, $\kappa$ is the grating strength, $\Delta n_{eff}$ is the difference in $n_{eff}$ between the areas with and without $Sb_2Se_3$, and $L$ is the length of the grating. For the case where the grating is sufficiently long relative to the grating strength (i.e., $\kappa \gg \pi/L$), the bandwidth simplifies to:

(2) $$\Delta\lambda \approx \lambda_B\left(\frac{2\Delta n_{eff}}{\pi n_g}\right)$$

From the above expression, we can see that $\Delta\lambda$ is directly proportional to $\Delta n_{eff}$. Thus, the reduction of $\Delta\lambda = 39.55$ nm in the embedded case (Fig. 2a) to $\Delta\lambda = 20.23$ nm in the evanescent case (Fig. 2c) is due to the ~2× decrease in $\Delta n_{eff}$ which can be seen when comparing the $n_{eff}$ plots in Fig. 1c-d. A weaker grating strength for a fixed grating length will also reduce the peak reflectivity at $\lambda_B$, which is equal to $R_{peak} = \tanh^2(\kappa L)$, as seen in the case of the evanescently coupled design. To account for this offset, the evanescent Bragg filter devices were fabricated instead with twice the number of periods as that of the embedded devices.

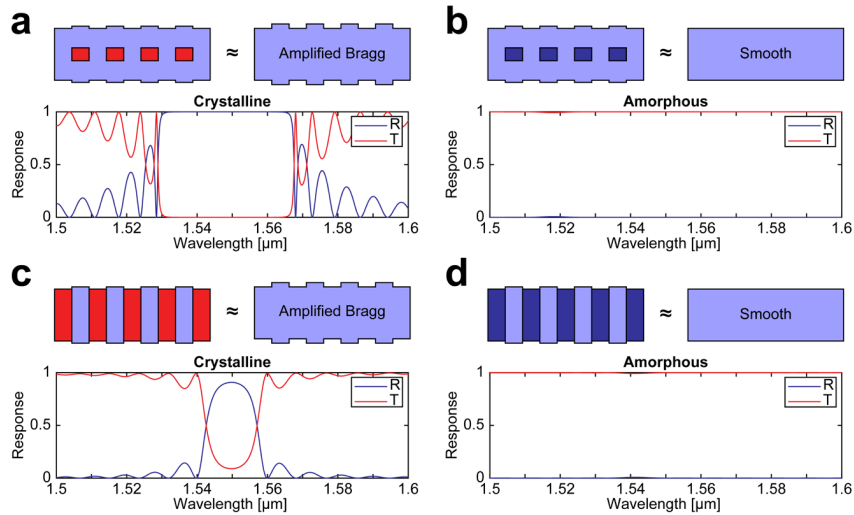

Fig. 2. (a)-(b) Simulated spectra of a Bragg grating with crystalline (a) and amorphous $Sb_2Se_3$ (b) embedded in the silicon waveguide grating. When the device is in the amorphous state as shown in (b), the transmission across the entire C-band and L-band is almost unity due to the precise matching of the effective refractive indices in the periodic structure. (c)-(d) Simulated spectra for the case of evanescently-coupled $Sb_2Se_3$ deposited on top of the waveguide for the crystalline (c) and amorphous phases (d). All simulations used TMM to model the spectra and accounted for the wavelength dispersion of the refractive index for both silicon and $Sb_2Se_3$. The number of periods for all simulations was held constant at N = 100.

*Modeling phase-shifted Bragg gratings with $Sb_2Se_3$*

An FP cavity can be created by placing a phase-shift inducing defect with an optical path length equal to an odd number of half periods within a Bragg grating. The length of the defect without any PCM added can be written in terms of an integer number of odd periods:

$$L_{defect} = \Lambda\left(m + \frac{1}{2}\right), \text{ where } m = 0, 1, 2, ... \tag{3}$$

where $\Lambda$ is the period of the grating and $m$ is a whole number. When a low-loss PCM is placed on the defect region, transitions between amorphous and crystalline phases will induce a resonance shift due to a change in $n_{eff}$ as explored by other groups in previous works [16], [17]. In addition to tuning the resonance position, we can also fully shift the resonance out of the stopband of the cavity upon a phase transition, thus removing the defect entirely. For this effect to happen, the change in $n_{eff}$ for the waveguide in the defect region must result in an odd multiple of π/2 phase shift for one of the phases while also providing an even multiple of π/2 phase shift in the other phase. Therefore, the length of the cavity can be designed at the first length to match the following condition:

$$L_{defect} = \frac{m\lambda_B}{2n_{eff,p1}} = \frac{(m+\frac{1}{2})\lambda_B}{2n_{eff,p2}}, \text{ where } m = 0, 1, 2, ... \tag{4}$$

where $n_{eff,p1}$ is the $n_{eff}$ of the FP defect waveguide in one of the PCM phases and $n_{eff,p2}$ the other. In order to center the resonance within the stopband, we can adjust the location of the stopband or the $n_{eff}$ of the FP cavity waveguide. This means the grating period ($\Lambda$), average effective index of the grating ($\bar{n}_{eff}$), and the width of the FP cavity are the critical dimensions which must be chosen to satisfy the above condition. However, if the length of the defect is too long, multiple FP resonances will be present for both the amorphous and crystalline state since the free spectral range of the cavity will be smaller than the bandwidth of the stopband. This effect can be seen when comparing the evanescently-coupled FP design in Fig. 3c-d with that of the shorter embedded design in Fig. 3a-b. This constraint necessitates a low-loss PCM with a large change in refractive index. We have previously demonstrated that only a ~11 μm length of Sb$_2$Se$_3$ can reversibly induce a π phase shift when deposited on silicon waveguides with integrated microheaters [20], making Sb$_2$Se$_3$ an ideal candidate for inducing transparency within the stopband of a phase-shifted Bragg grating.

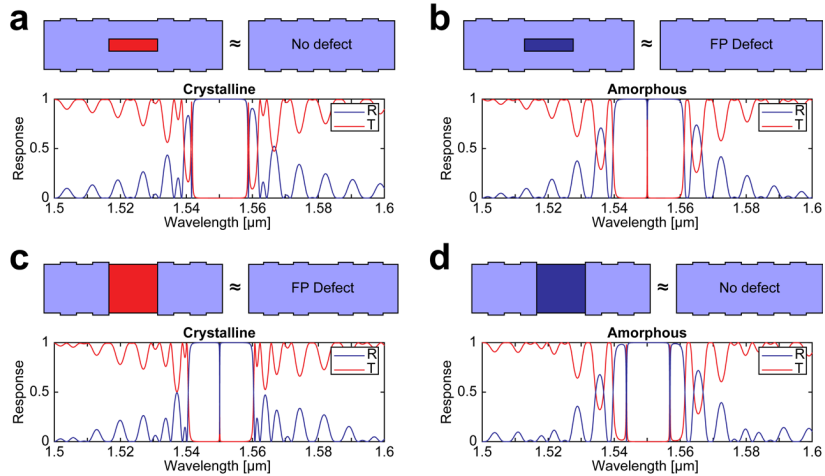

Fig. 3. (a)-(b) Simulated spectra of a phase-shifted Bragg grating with crystalline (a) and amorphous Sb$_2$Se$_3$ (b) embedded in the phase-shifted defect section. The length of the defect is chosen such that Δn$_{eff}$ of the defect is equal to π when Sb$_2$Se$_3$ is switched between its amorphous and crystalline phases. (c)-(d) Spectra for a phase-shifted Bragg grating for an evanescently coupled design. The presence of a Fabry-Perot (FP) resonance in stopband when Sb$_2$Se$_3$ is in the crystalline phase is determined by whether the defect is equivalent to either an even or odd number of grating periods.

## 4. Fabrication and Experimental Measurements

Devices were fabricated according to the designs shown in Fig. 2 and Fig. 3. Fifty variations on the Bragg design and two hundred fifty variations of the FP Cavity devices were fabricated to account for fabrication non-idealities. These variations included average waveguide width in the Bragg mirrors, width differences between segments, the total number of periods, and the length of the FP cavity in the FP cavity devices. Note that the embedded Bragg filters were tested with 100 and 200 periods while the evanescent Bragg filters were tested with 200 and 400 periods to account for the reduced contrast $\Delta n_{eff}$ of the latter design.

The proposed designs were patterned using an ELS-G100 electron-beam lithography system on a silicon-on-insulator (SOI) platform (220 nm Si on 3 μm buried oxide from University Wafer) using ZEP positive resist. Reactive ion etching (RIE) in $SF_6/C_4F_8$ was then carried out to etch away 220 nm of Si. A 30 nm thin film of $Sb_2Se_3$ was thermally evaporated (120-nm-thick $Sb_2Se_3$ in the case of the embedded design) and a second electron-beam lithography step was used to pattern the $Sb_2Se_3$ layer using MaN-2403. The unexposed regions are subsequently etched away using RIE in $CF_4$ forming the $Sb_2Se_3$ patches on top of the waveguide and finally everything was capped using 10 nm of sputtered $SiO_2$ to avoid oxidation. A Santec TSL-570 tunable laser was used as the laser source and optical signals were collected by a Santec MPM-210 photodetector to capture transmission spectra of the Bragg devices.

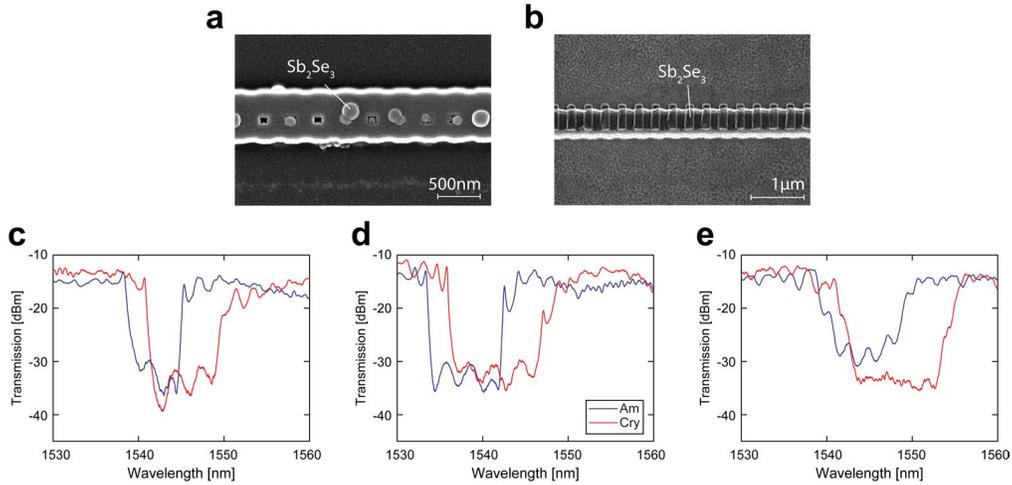

Fig. 4. Results of fabricated Bragg devices. (a)-(b) SEM Images of fabricated devices. Devices with $Sb_2Se_3$ embedded (a) and evanescently coupled (b) on waveguide segments of a Bragg grating. The poor quality of the filling and nonuniformity of the $Sb_2Se_3$ in the waveguide resulted in minimal spectral shift after crystallization. (c)-(e) Spectra of devices with $Sb_2Se_3$ deposited on top of the waveguide. The device in (b) shows an example of poor alignment between the $Sb_2Se_3$ and Bragg grating which reduces effectiveness of the refractive index contrast after phase transition. Multiple devices exhibit a red-shift in Bragg wavelength and increase in bandwidth upon crystallization.

Fig. 4a-b show SEM images of the fabricated devices. Both designs exhibit misalignment between the two electron-beam lithography steps and lithography smoothing, which affected all device types. The embedded PCM devices (see Fig. 4a) exhibit an incomplete and nonuniform filling of the trenches by the phase change material and passivation layer. As a result, the embedded PCM devices did not operate according to their design. The evanescent PCM devices displayed unintentional alignment offsets between the $Sb_2Se_3$ and photonic device layers (Fig. 4b) which reduced functionality. Fig. 4c-e show different measured spectra across three evanescent PCM devices, which still displayed an expected increase in $\bar{n}_{eff}$ of the grating to produce a $\lambda_B$ red shifting of approximately 5 nm. The red shift, in turn, leads to extinction ratios of ~20 dB for the wavelengths outside the overlapping stopbands. Upon

crystallization, Fig. 4c-e also exhibit broadening of the stopband bandwidth, which corresponds to an expected increase in $\Delta n_{eff}$ between segments. Fig. 4c exhibits a response for a device with only 200 periods while the devices in Fig. 4d-e had 400 periods. The uniform response versus number of periods can be attributed to accumulation of phase error in the Bragg mirrors.

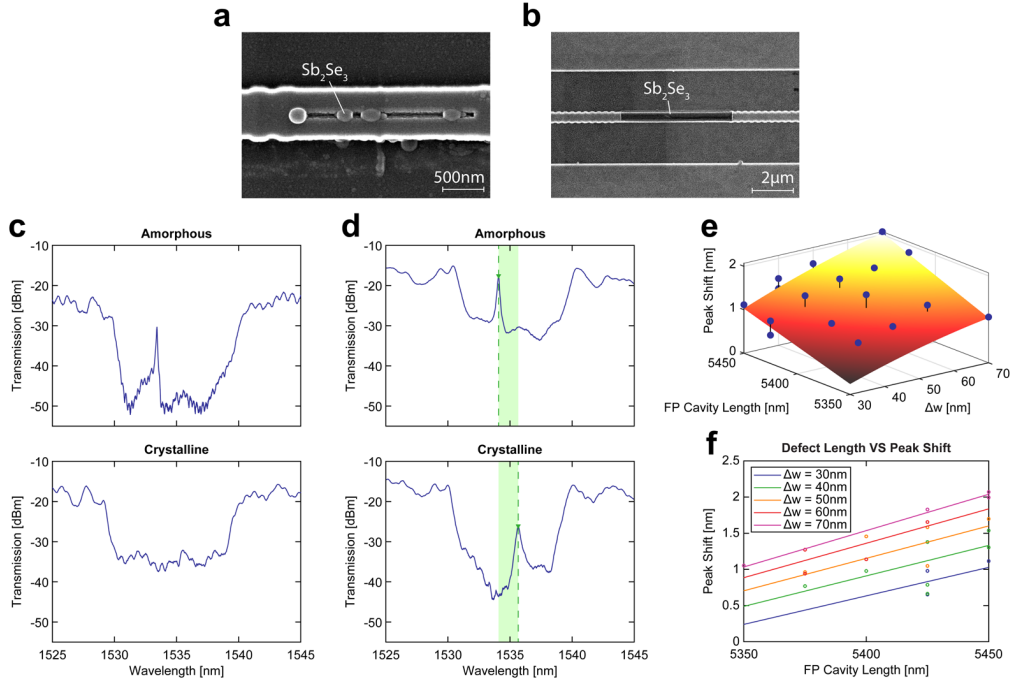

Fig. 5. Results of fabricated Fabry-Perot Etalon devices. (a)-(b) SEM Images of fabricated devices. Devices with $Sb_2Se_3$ embedded (a) and evanescently coupled (b) on the defect segment of a FP device. The poor quality of the filling of the $Sb_2Se_3$ in the embedded waveguide resulted in minimal effects after crystallization. The phase-shifted device shown in (b) is much less sensitive to alignment compared to the Bragg grating design (Fig. 4a). (c)-(d) Spectra of devices with $Sb_2Se_3$ deposited on top of the defect. The device in (c) shows an example of expected device behavior. Device (d) exhibits the characteristic red shift of the passband. Passband peak shift was shown to be related to both FP cavity length and $\Delta w$ of the Bragg reflectors and is shown in (e)-(f).

Fig. 5a-b show SEM images of the fabricated Fabry-Perot Etalon (FP) devices. As with the Bragg filter devices, both designs exhibit fabrication imperfections, though the evanescent FP design is much more tolerant to small misalignments than the Bragg filter due to the relative size of the $Sb_2Se_3$ area compared to the alignment accuracy. Fig. 5a shows an embedded PCM device with an incomplete filling of the trench by the phase change material and the passivation layer, which similarly led to devices unresponsive to thermal annealing. In contrast, some misalignment of the PCM over the FP cavity in the evanescent devices did not lead to major changes in expected performance. Fig. 5c shows an effective device which shifts the passband out of the stopband's bandwidth upon crystallization, and Fig. 5d exhibits a device with a 1.583 nm shifting of the passband upon crystallization with extinction ratios of exceeding 25 dB. The device in Fig. 5c used an FP defect that was designed to be 5.425 μm long on a 470 nm wide waveguide. The passband of the device in Fig. 5c exhibits a Q factor of about $1.6 \times 10^4$ while the device in Fig. 5d exhibits Q factors of $8.3 \times 10^3$ and $4.0 \times 10^3$ in the amorphous and crystalline phases respectively. The shift in resonance wavelength of the FP cavity upon phase transition is independent of the average width or number of periods of the Bragg reflectors (minimum number of periods was 100), but is proportional to changes in the FP defect length and modulation strength of $\Delta n_{eff}$ in the width-modulated passive Bragg mirrors (denoted as

$\Delta w$ in Fig. 5f). The increase in red shift with respect to $\Delta n_{eff}$ (which is proportional to $\Delta w$) between segments agrees with TMM simulations where FP length and $\bar{n}_{eff}$ are held constant for the given geometries that are shown in Fig. 5. The experimental passband red shift was measured in 23 devices, as plotted in Fig. 5e and fitted as a function of FP cavity length (linearly) and to the width difference between segments in the Bragg mirrors (quadratically). Devices which exhibited larger shifts, such as the device in Fig. 5d, did exist but were not plotted in Fig. 5e since the peak was unable to be measured within the stopband region. The derived contour lines for $\Delta w$ are shown in Fig. 5f.

## 5. CONCLUSION

We have proposed and experimentally demonstrated nonvolatile switchable devices using Bragg gratings with various designs and functionalized with $Sb_2Se_3$. Our experimental results demonstrate the feasibility of the different tunable designs simulated in Section 3. Namely, Bragg grating devices with stopband shifts of ~5 nm and extinction rations of 20 dB upon phase transition, and Fabry-Perot etalon devices whose resonances can be shifted as a function of geometrical parameter of the $Sb_2Se_3$ cell leading to extinction rations of ~20 dB. Moreover, we explored both embedded and evanescently coupled PCM devices; however, the former, while designed to display the strongest modulation, were highly challenging to fabricate and require further refinement. Finally, the devices we propose in this work can be readily integrated into photonic integrated circuits featuring doped-silicon microheaters for electrically controlled reversible switching [18], [20]–[23]. Our results pave the way towards zero-static power reconfigurable Bragg gratings for high-performance filtering and routing in photonic integrated circuits.

**Funding.** This work was supported in part by the U.S. National Science Foundation under Grants ECCS-2028624, DMR-2003325, ECCS-1901864, ECCS-2210168/2210169, ECCS-2132929 as well as by the Office of Naval Research (ONR award #N000141410765). N.Y. acknowledges support from the University of Pittsburgh Momentum Fund. C.R acknowledges support from the Minta Martin Foundation through the University of Maryland.

**Disclosures.** The authors declare no conflicts of interest.

**Data availability.** Data underlying the results presented in this paper are not publicly available at this time but may be obtained from the authors upon reasonable request.